# HPR-Mul: An Area and Energy-Efficient High-Precision Redundancy Multiplier by Approximate Computing

Jafar Vafaei, Omid Akbari

*Abstract*— For critical applications that require a higher level of reliability, the Triple Modular Redundancy (TMR) scheme is usually employed to implement fault-tolerant arithmetic units. However, this method imposes a significant area and power/energy overhead. Also, the majority-based voter in the typical TMR designs is highly sensitive to soft errors and the design diversity of the triplicated module, which may result in an error for a small difference between the output of the TMR modules. However, a wide range of applications deployed in critical systems are inherently error-resilient, i.e., they can tolerate some inexact results at their output while having a given level of reliability. In this paper, we propose a High Precision Redundancy Multiplier (HPR-Mul) that relies on the principles of approximate computing to achieve higher energy efficiency and lower area, as well as resolve the aforementioned challenges of the typical TMR schemes, while retaining the required level of reliability. The HPR-Mul is composed of full precision (FP) and two reduced precision (RP) multipliers, along with a simple voter to determine the output. Unlike the state-of-the-art Reduced Precision Redundancy multipliers (RPR-Mul) that require a complex voter, the voter of the proposed HPR-Mul is designed based on mathematical formulas resulting in a simpler structure. Furthermore, we use the intermediate signals of the FP multiplier as the inputs of the RP multipliers, which significantly enhance the accuracy of the HPR-Mul. The efficiency of the proposed HPR-Mul is evaluated in a 15-nm FinFET technology, where the results show up to 70% and 69% lower power consumption and area, respectively, compared to the typical TMR-based multipliers. Also, the HPR-Mul outperforms the state-of-the-art RPR-Mul by achieving up to 84% higher soft error tolerance. Moreover, by employing the HPR-Mul in different image processing applications, up to 13% higher output image quality is achieved in comparison with the state-of-the-art RPR multipliers.

## I. Introduction

**M**ultiplier is one of the main arithmetic blocks in different digital systems and its design metrics (i.e., delay, area, and power/energy consumption) can significantly impact the system performance. Digital signal processing (DSP), deep neural networks (DNNs), matrix multiplication (MM), and stochastic computations are examples of these systems, in which the efficiency of the multipliers has a vital role [1]. Moreover, in critical applications where achieving a given level of reliability is essential, multipliers are considered important parts that can be hardened against fault by employing various redundancy schemes, such as Triple Modular Redundancy (TMR) [2], residue code [3], and parity prediction [4] methods.

In the TMR designs, the outputs of the replicated modules are forwarded to a majority-based voter, in which to determine the valid output by the voter, at least two modules have to provide the correct output. Also, the replicated modules may be implemented in different ways but with the same functionality to decrease the impact and possibility of common-mode failures (CMFs) [5]. However, the design diversity of the replicated modules may result in some small differences in the modules' outputs. Furthermore, various types of errors induced by the different factors such as electromagnetic inference (EMI), manufacturing process defects, aging, radiations, and the Single-event effects (SEEs) that are more likely to happen in the nanoscale era, can affect the normal operation of the replicated modules. SEEs are the disruptions to the normal operation of a device caused by a single energetic particle striking a sensitive area in the device, which can be classified into different categories such as Single Event Transients (SETs) and Single Event Upsets (SEUs) [6].

SETs are temporary glitches or errors in electronic devices when a single particle strikes a sensitive node, causing a momentary disruption in the device's operation. SETs do not result in permanent damage to the device, but they can cause transient errors that may affect the device's functionality. SEUs are changes in the device's state or data. When a single energetic particle strikes a sensitive node in the device, it can flip a bit in a memory cell or register, leading to an error in the device's operation. SEUs can have significant consequences, especially in critical applications where data integrity is essential. Overall, SEEs are a critical consideration in fault-tolerant design, as they can have implications for the reliability and safety of electronic systems. Therefore, designers implement various strategies to mitigate the impact of SEEs, such as error detection and correction mechanisms, shielding, redundancy, and error recovery techniques. The aforementioned error factors can make some small differences in the outputs of the replicated modules, which can be considered as an error by the typical TMR voters [5].

Reduced precision redundancy (RPR) is one of the techniques that can resolve these challenges, as well as decrease the power/energy and area overheads of the TMR designs [7][8].

Jafar Vafaei is with the School of Electrical and Computer Engineering, University of Tehran, Tehran 14395-515, Iran (e-mail: jafar.vafaei@ut.ac.ir).

O. Akbari (corresponding author) is with the Department of Electrical and Computer Engineering, Tarbiat Modares University, Tehran 14115-111, Iran (e-mail: o.akbari@modares.ac.ir).

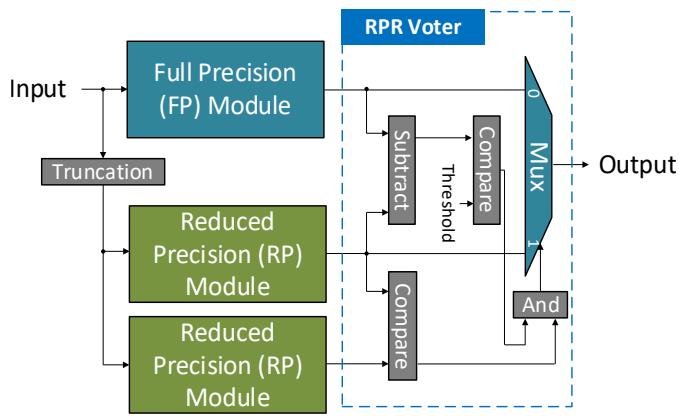

Fig. 1. Generic architecture of a typical reduced precision redundancy (RPR) module [9].

The generic architecture of an RPR design is shown in Fig. 1 [9], which is composed of one full precision (FP) and two reduced precision modules (e.g., adder [10] or multiplier [11]), along with an RPR voter. Note that the reduced precision modules are usually obtained by the truncation method, in which some least significant bits (LSBs) of the modules are removed. As shown in Fig. 1, the RPR voter is composed of several extra modules, including a subtractor, comparator, logical gates, and a multiplexer, which can increase the area and delay overheads, depending on the bit width of the modules' outputs.

As an example, in the state-of-the-art RPR multiplier presented in [11], the FP multiplier (i.e., an $N$-bit multiplier) produces the full precision result, and the RP multipliers (i.e., $N$-$M$ bit multipliers) generate the reduced precision outputs, where the $M$ LSB bits of the RP multipliers were truncated. The subtractor in the RPR voter measures the difference between the FP multiplier output and one of the RP multipliers, and when this difference exceeds a predefined threshold, an error is presumed to have occurred. In this case, if the comparator at the output of the RP multipliers shows a difference, it is assumed the error has occurred in the RP multipliers, and the output of the FP multiplier is selected by the multiplexer to be forwarded to output. Otherwise, the error has occurred in the FP multiplier, and the output of one of the RP multipliers is selected by the multiplexer.

In recent years, approximate computing has been of great interest since it employs the error resiliency of applications to achieve power/energy consumption and area improvements at the cost of inducing some errors at the output. Multimedia processing, digital signal processing, data mining, computer vision, machine learning, and big data analysis, are some examples of error-resilient applications, in which a span of outputs near to the exact (golden) one is acceptable [5]. As an example, in an image processing application, the output images may be perceived by the human eyes that are not sensitive to small inaccuracies. However, in the error-resilient applications deployed in critical systems, achieving a given level of reliability has more priority compared to generating the exact outputs. Therefore, in these cases, the approximate computing paradigm can be utilized for gaining the aforementioned advantages, while still providing the required level of reliability [5][12].

In this paper, we propose a high precision redundancy multiplier (HPR-Mul) that exploits approximate computing to achieve significant power/energy and area improvements, while retaining the required level of reliability. Specifically, we design the HPR voter based on mathematical formulas to achieve a simple voter compared to the RPR voter shown in Fig. 1. Furthermore, we use the intermediate signals of the FP multiplier as the inputs of the RP multipliers to increase their accuracy against soft errors. The novel contributions of this paper in a nutshell are:

1- Proposing a generic and systematic method to design high precision redundancy (HPR) multipliers, while retaining the required level of reliability (Section III).

2- Achieving a high accuracy and fault tolerance by enhancing the accuracy of RP multipliers through exploiting the intermediates signals of the FP multiplier (Subsection III.B).

3- A mathematical method to design the voter of HPR-Mul to achieve a simpler voter compared to the state-of-the-art RPR multipliers (Subsection III.C).

4- Investigating the soft error tolerance of the proposed HPR-Mul, and comparing the results with typical TMR and state-of-the-art RPR multipliers (Subsection IV.A).

5- Extracting the design metrics (power, delay, and area) of the HPR-Mul, and comparing them with those of the typical TMR multipliers, as well as the state-of-the-art RPR multipliers (Subsection IV.B).

6- Assessing the efficacy of the studied redundancy schemes in real-world image processing applications (Subsection IV.C).

**Paper Organization:** In Section II, prior related works on reduced precision redundancy-based are reviewed. Then, the architecture of the proposed HPR-Mul is presented in Section III. Section IV deals with the simulation results and the effectiveness evaluation of the studied redundancy schemes. Finally, the paper is concluded in Section V.

## II. RELATED WORK

In this section, different efforts dealing with approximate computing in reliable systems are studied. Specifically, we review key research focused on reduced precision redundancy methods are reviewed.

A comprehensive review of the different techniques of approximate computing is found in [13], where these techniques are classified into different system layers, ranging from hardware and architecture to software layers. Also, exhaustive surveys on approximate computing and its fault tolerance have been conducted in [14]-[16].

In [17], an inexact double modular redundancy (IDMR) voter was proposed, in which, when the difference between the outputs in the DMR system is higher than a predefined threshold, an error signal is generated. Otherwise, the average of outputs is computed as the correct output. The proposed method of IDMR was extended to TMR voters in [18]. An approximate TMR voter for loop-based algorithms was proposed in [19], where by varying the number of loop iterations, an accuracy-performance tradeoff is achieved. In [20], the authors have addressed the challenges of selective hardening in the arithmetic circuits by exploring the different tradeoffs between reliability and cost. In [21], a Boolean factorization-based method was proposed to compose approximate logics for redundant systems and reduce the area overheads. In [22], a partial TMR-based design was proposed for the FPGAs that employ approximate logics in the redundant modules. A transistor-level voter hardening method was proposed in [23], that analyzed the voter inputs and constructed a quadded transistor-based redundancy voter for masking the transient faults.

Different schemes were employed to protect the arithmetic units against faults. In [3], the modulo of output is checked to detect the error occurrence, while in [4], the parity method is used as an error detection knob. In both of these works, two replicated modules were employed to correct the output, when an error occurred. However, these methods impose a considerable area cost.

One of the promising techniques to reduce the overheads of the $N$ Modular Redundancy (NMR) based techniques is using $(N-1)$ reduced precision redundant modules along with a main full precision module, which is known as the reduced precision redundancy (RPR) scheme. Because of the significant area and power/energy consumption reduction by the reduced precision modules, the RPR scheme can achieve significant improvements. As an example, in [24], it has been shown that the failure rate of an RPR design was improved by 200× against Single Event Upsets (SEUs), whereas its area was reduced by 50% compared to the typical TMR scheme. However, the overheads of the complex voter in the RPR schemes, induced by error detection and correction are still considerable [25]. An RPR voter for the carry propagate adders (CPAs) was proposed in [10]. This work splits the structure of the main adder (main module in a TMR design) into two higher and lower parts, and only replicates the higher part as the redundant modules of the TMR design, where the output carry of the main adder lower part is used as the input carry of the replicated adders. Therefore, the required voter is simplified and its cost is reduced.

Loop perforation is one of the well-known software-level approximate computing techniques, in which, the number of iterations in "for" loops is modified manually. In [26], loop perforation was used to design an approximate software-based fault-tolerant system to diminish the execution time overheads. In [27], a duplication with comparison (DWC)-based fault tolerant scheme for image processing applications was proposed, in which one of the replicated modules was replaced with an approximated one. Then, this design is used to distinguish between usable and unusable images by a convolutional neural network (CNN). In [28], approximate computing methods are used along with radiation-induced mitigation techniques to reduce the overheads of fault-tolerant systems. [32] used logic optimization techniques to find an appropriate combination of three approximate modules to compose an approximate TMR design.

As reviewed in this section, most state-of-the-art RPR works used expensive voters and almost were limited to adders. However, the multipliers use significantly more area and power/energy in digital systems. In this paper, we propose a systematic method to design a high precision redundancy multiplier (HPR-Mul), which is based on simplifying the voter design based on mathematical formulas, as well as enhancing the accuracy and error tolerance of the design by exploiting some intermediate signals of FP multiplier as the input of RP multipliers.

## III. PROPOSED HIGH PRECISION REDUNDANCY MULTIPLIER

In this section, first, the implementation of a conventional block-level multiplier is discussed. Then, the studied block level multiplier is used to compose the proposed high-precision redundancy multiplier (HPR-Mul). Afterward, a mathematical method to design the voter of the HPR-Mul is introduced.

### A. Conventional Block-Level Multipliers

A multiplier can be implemented by using smaller blocks [1]. This design along with a numerical example is shown in Fig. 2, where the two inputs of $A$ and $B$ are split into two higher ($H$) and lower ($L$) parts. Then, the main multiplier is composed of four smaller blocks, where $A_H$ ($B_H$) and $A_L$ ($B_L$) are the upper and lower parts of the input $A$ ($B$), respectively. Note that for an input with the size of $N$, the size of the upper and lower parts is ($N$-$K$) and $K$, respectively. We leverage this structure to compose the FP multiplier of the proposed HPR-Mul, and then, employ its intermediate signals to increase the accuracy of RP multipliers, to achieve higher accuracy in faulty conditions compared to the typical TMR multipliers and RPR designs.

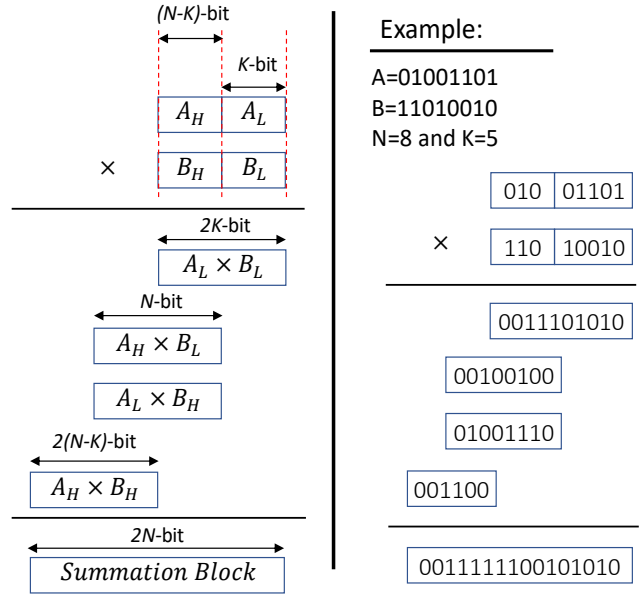

Fig. 2. Composing a larger multiplier with smaller multiplier blocks, along with an example to show the functionality of this method.

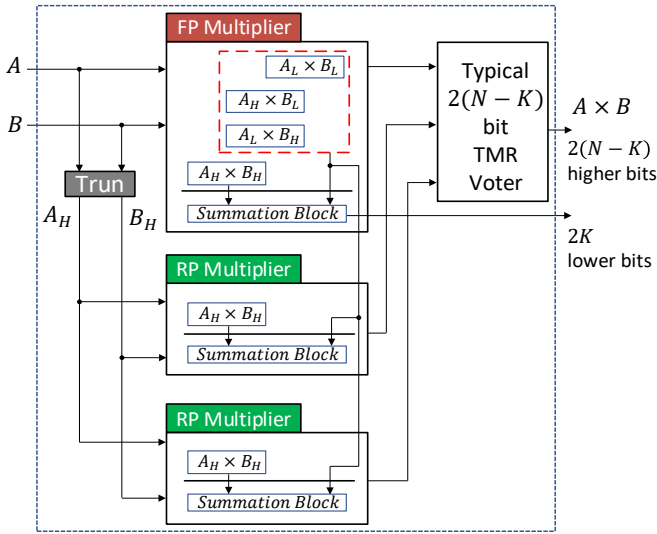

Fig. 3. Block Diagram of the HPR-MUL

*B. Proposed HPR-MUL*

As shown in Fig. 1, in an RPR multiplier, there is not any dependency between the main (i.e., the FP multiplier) and redundant (i.e., the RP) multipliers. In our proposed HPR-MUL, we create a dependency between these multipliers to simultaneously increase its accuracy and fault tolerance feature with a slight overhead. Fig. 3 shows the structure of the proposed HPR-MUL.

To compose the HPR-Mul, at first, we implement the main multiplier in the block level form (see Fig. 2) consisting of four smaller multipliers, where their outputs are forwarded to the summation block for computing the final result. Also, the redundant multipliers are implemented with a smaller size that corresponds to the $a_H \times b_H$ block in Fig. 2. Note that the RP modules can be obtained by the truncation method, e.g., the $K$ lower bits of the inputs of an *N*-bit multiplier are truncated, and therefore, *(N-K)*-bit multipliers are used instead of full precision ones, where the output of these multipliers is *2(N-K)*-bit. This method can significantly reduce the area and power/energy consumption of the truncated module compared to the FP module. Now, the intermediate results calculated by the $a_L \times b_L$, $a_H \times b_L$, and $a_L \times b_H$ blocks of the FP multiplier (red-colored dashed box in Fig. 3) are forwarded to the summation blocks of the RP multipliers, to enhance their accuracy.

Note that, depending on the size of the smaller blocks a tradeoff between the accuracy and power/area of the HPR-Mul is achievable. In particular, the size of the $a_H \times b_H$ block is determinant since it is replicated in the redundant multipliers, and thus, various levels of area and power/energy savings are achievable. Next, the outputs of the FP multiplier and RP multipliers are forwarded to a simple typical *2(N-K)*-bit TMR voter, instead of expensive voters used in typical RPR designs (see Fig. 1), since the computed lower *2K*-bit of the FP multiplier and RP multipliers are the same. Therefore, depending on the *K*, the HPR voter may achieve lower design metrics (i.e., delay, power/energy consumption, and area) compared to an *N*-bit typical TMR voter. It's worth noting that the proposed HPR-Mul retains the reliability, i.e., still the proposed design is triplicated in implementation, where only some LSB bits of the reduced precision (RP) multipliers are relaxed. Moreover, employing the approximate computing in TMR multipliers may resolve the challenge of typical TMR multipliers and increase the system dependability.

Details of implementing the HPR-Mul are depicted in Fig. 4, in which all the components and their interconnections are depicted. Moreover, the required values for the RP Multipliers are depicted with a numerical example on the right side of this figure. As shown in this figure, the RP multipliers are composed only of a $a_H \times b_H$ multiplier and a $2(n - k)$-bit adder.

It's worth noting that similar to some state-of-the-art RPR works (such as [10]), in our proposed HPR multiplier, we created dependency between the FP and RP modules, and achieved a simpler voter design, to obtain a moderate fault tolerance feature with a considerable reduced overhead. Due to this dependency, the soft error effects may be propagated from the lower part of the FP multiplier into the RP multipliers. However, the weight of the $A_H B_H$ block in the accuracy of the output of the summation block is significantly higher than the lower parts. Based on the results that will be presented in Subsection IV.A, in our soft error tolerance examinations, we injected multiple faults in different parts of all FP and RP multipliers, simultaneously, where the results show a significantly higher soft error tolerance compared to the state-of-the-art RPR schemes.

In the next subsection, a mathematical method will be proposed to determine the parameter *K* of the HPR-Mul, such that the accuracy of the HPR-Mul is retained in the acceptable range, without requiring any extra hardware such as those used in RPR voters.

*C. Selecting Parameter K for the proposed HPR Multiplier*

In this subsection, we propose a mathematical method to select a proper value for *K*, such that the obtained configuration of the proposed HPR-Mul meets the user-defined quality degradation upper bound ($Q_{DUB}$).

Note that, the quality degradation upper bound of the HPR-Mul can be obtained from the higher level of abstraction (e.g., the application layer), in which a number of the proposed HPR-Mul may be deployed. In such a case, a quality translation (quality mapping) step, similar to those proposed in [29] and [31] is required. In [29], the adder and multiplier inside the ALU of all processing elements (PEs) of a coarse-grained reconfigurable architecture (CGRA) were replaced with the quality adjustable adder and multiplier proposed in [30] and [34], respectively, which result in approximate CGRA with quality scalable processing elements (QSPEs) that provide the ability to implement an application with various quality levels. To exploit this feature, a heuristic technique was proposed that is applied to the data flow graph (DFG) of a given application, and based on an integer linear programming (ILP) method, translates the desired quality level of the application into the approximate mode (quality level) of the QSPEs. Note that the desired quality level of the application can be obtained from a

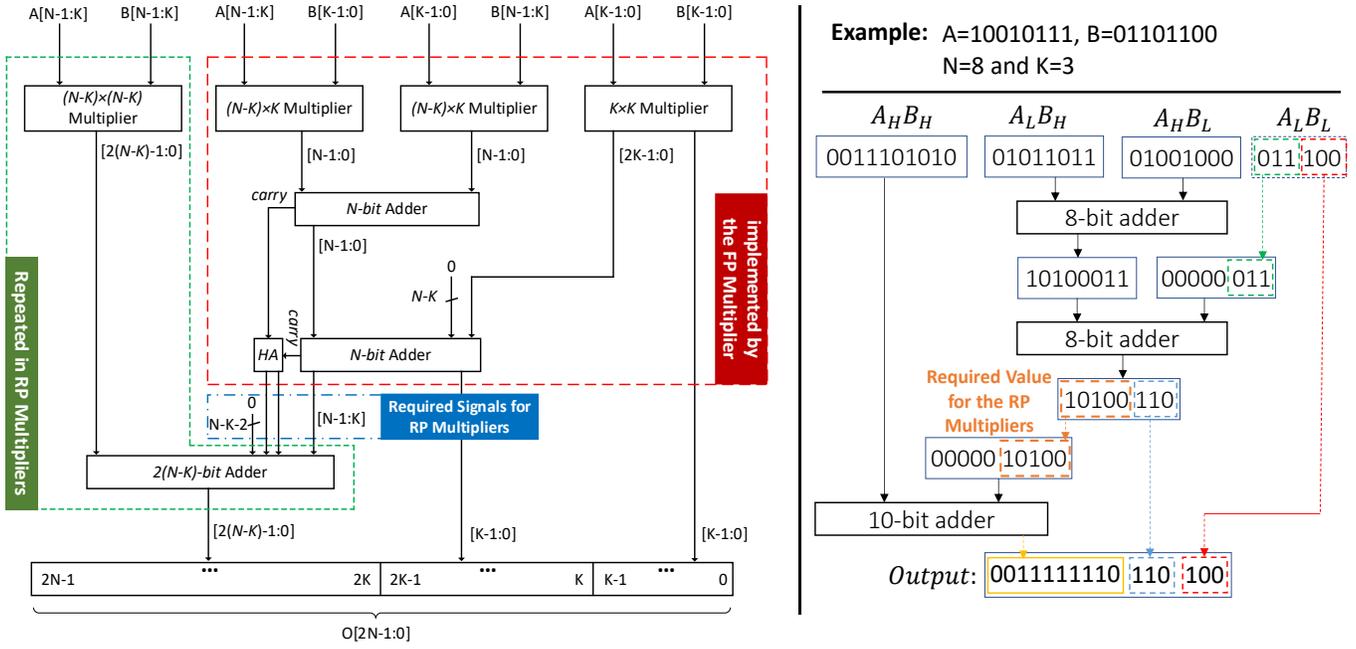

Fig. 4. The implementation of the Full Precision (FP) multiplier and required signals for the Reduced Precision (RP) multipliers, along with a numerical example to show its functionality.

Pareto optimal curve, in which the different quality levels of the target application correspond to various approximate modes of the QSPEs. Thus, through this process, from a desired quality (energy consumption) level of the application, the quality level of all adders and multipliers inside the QSPEs is determined. In [31], an approximate architecture composed of quality adjustable vector processors was proposed, where a quality control unit translates the instruction-level quality specifications to hardware accuracy for scaling the precision of the operations.

Now, to select a proper value for $K$ while meeting the $Q_{DUB}$, an error metric is required to show the amount of the output inexactness induced by relaxing the $K$ lower bits of the RP multipliers in the HPR-Mul. In the approximate computing literature, different error metrics have been proposed, such as Error Rate (ER), Error Distance (ED), Mean Error Distance (MED), Mean Relative Error Distance (MRED), and Mean Normalized Error Distance (MNED), where their formula have been shown in TABLE I. As shown in this table, most of the error metrics are defined based on the ED, which is the difference of the exact and approximate (inexact) output values. We also employ the ED metric to calculate the parameter $K$.

Fig. 5 shows the steps of calculating the parameter $K$. First, by receiving the $N$ (i.e., the bit width of the multiplier inputs) and $Q_{DUB}$, the Maximum Tolerable Error Distance (MTED) metric is calculated by [5]:

$$MTED = (2^N - 1) \times \frac{Q_{DUB}}{100} \quad (1)$$

Now, the MTED is rounded down to the nearest power of two in the form of $2^{2m}$, where $m$ is an integer value. This rounding down ensures that the $Q_{DUB}$ is always met [5].

TABLE I  SOME ERROR METRICS DEFINED FOR EVALUATING THE ACCURACY LOSS IN THE APPROXIMATE SYSTEMS

| Error Metric | Formulation | Description |
|---|---|---|
| ER | $\frac{\#errorneous\ outputs}{T}$ | $T$: Total number of output samples |
| ED | $\|O - O'\|$ | $O$: Exact Output $O'$: Aprx. output |
| MED | $\frac{1}{T}\sum_{i=1}^{T}\|ED_i\|$ | - |
| MRED | $\frac{1}{T}\sum_{i=1}^{T}\frac{\|ED_i\|}{O_i}$ | - |
| MNED | $\frac{MED}{D}$ | $D$: Maximum Possible value at the output |

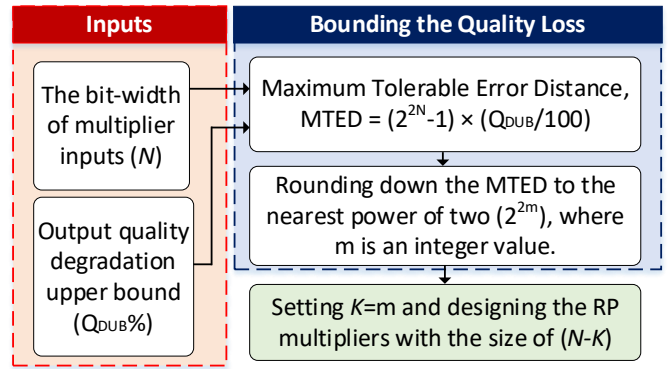

Fig. 5. Steps for calculating the $K$ based on the user-defined quality constraint (i.e., $Q_{DUB}$).

**Inputs:** *A*=151 (10010111), *B*=108 (01101100), *N*=8, $Q_{DUB}$=7%
**Outputs:** *MTED*=17.85 => K=2

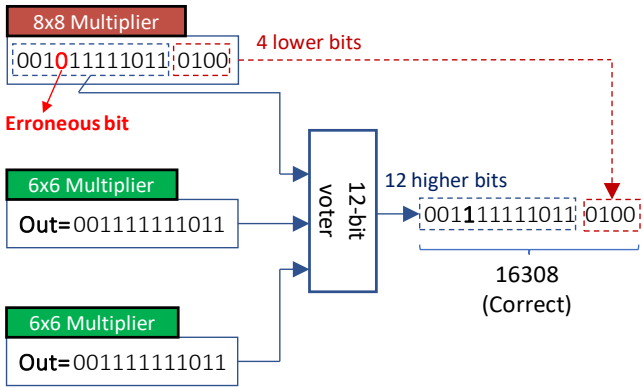

Fig. 6. A numerical example to show the functionality of the HPR-MUL in the presence of error.

Then, by setting $K = m$, the size of FP and RP multipliers is determined, i.e., the parameter *K* is obtained by:

$$K = \frac{1}{2}\lfloor \log_2 MTED \rfloor \quad (2)$$

Now, the *K* lower bits of the inputs of RP multipliers are relaxed, i.e., the TMR technique is not applied to these bits. In this case, the output of these multipliers is *2(N-K)*-bit, and thus, the size of the voter is also *2(N-K)*-bit (see Fig. 3). Thus, instead of a *2N*-bit voter used in typical TMR multipliers and the expensive voters used in typical RPR designs shown in Fig. 1, a simpler voter with lower design metrics (i.e., delay, power/energy consumption, and area) is required to compose the HPR-Mul.

To show the functionality and error tolerance of the proposed HPR-Mul, a numerical example of the proposed method is shown in Fig. 6. In this example, by assuming *N=8* and the $Q_{DUB} = 7\%$, using (2), the value of parameter *K* will be 2. Thus, 2-bit of the inputs of RP multipliers are truncated, and thus, the outputs of RP multipliers are 12-bit (i.e., *2(N-K)*-bit) instead of 16-bit (i.e., *2N*-bit). In this case, the size of the voter is also 12-bit.

## IV. RESULTS AND DISCUSSION

In this section, first, we evaluate the effectiveness of the proposed HPR-Mul against soft errors and compare it with the typical TMR-based multiplier, as well as the state-of-the-art RPR multiplier presented in [11]. Next, we investigate the design parameters (delay, power, and area) of these schemes. Finally, we examine the efficacy of the studied TMR schemes in the different image processing applications.

### A. Soft Error Tolerance

To investigate the effect of soft errors on the discussed redundancy schemes, we inject uniformly distributed random errors in their components. Fig. 7 shows the used noise model to inject errors, where inspired by [5], we proposed a controllable noise injection model, in which a noise source is placed at the input of multipliers. Next, depending on a predefined probability of the flip ($P_f$) factor and in a bitwise manner, each bit of the inputs of the multiplier is flipped independently. For this, corresponding to each bit, a uniform random value between 0 to 1 is generated, and if the generated value is less or equal to the $P_f$, the input bit of the multiplier is flipped. Otherwise, that bit remained unchanged, i.e., it was not affected by the noise. Thus, in the following examinations, multiple faults are injected on different multipliers with the factor of $P_f$. Note that for this evaluation, we used the same random inputs for all the studied redundancy schemes. Finally, the Mean Square Error (MSE) value of the output is calculated to show the soft error tolerance of the investigated redundancy schemes [11].

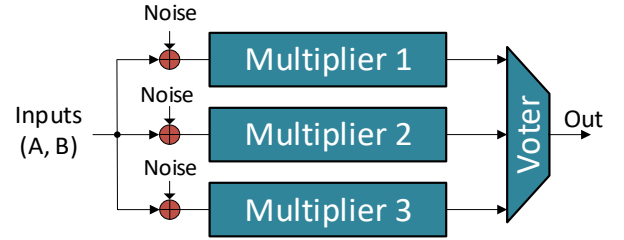

Fig. 7. The used noise model to insert errors at the inputs of multipliers.

Fig. 8 shows the MSE of RPR and HPR schemes, for $N = 8$ and the different values of *K*, including 2, 4, and 6, where the results are normalized to those of the typical TMR scheme. Based on the results, for the $P_f$ in the range of [0.001, 0.02] and for $k = 2$, 4, and 6, the proposed HPR schemes achieve, on average, 77%, 81%, and 84% lower MSE, respectively, compared to the state-of-the-art RPR multiplier.

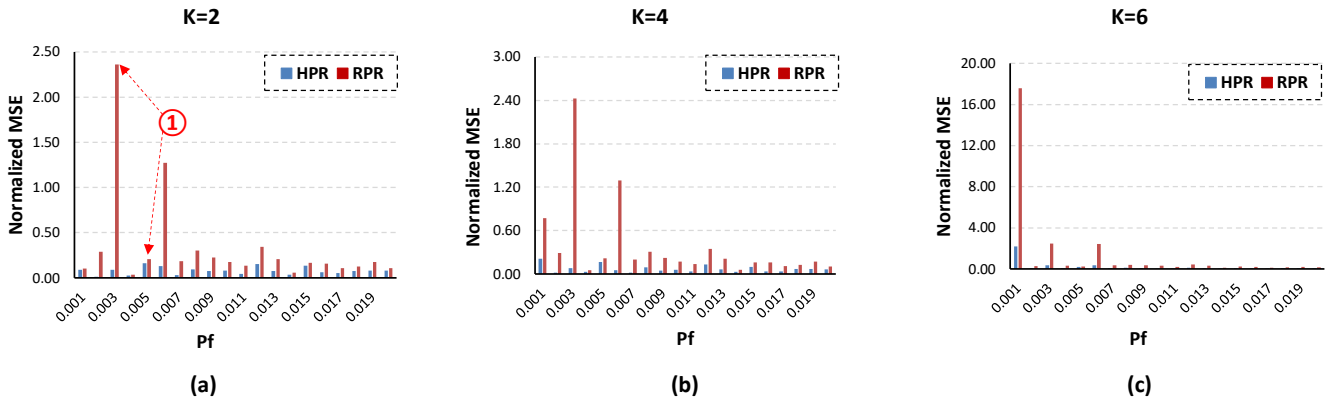

Fig. 8. MSE of the HPR and RPR multipliers normalized to the typical TMR multipliers, using the noise model presented in Fig. 7.

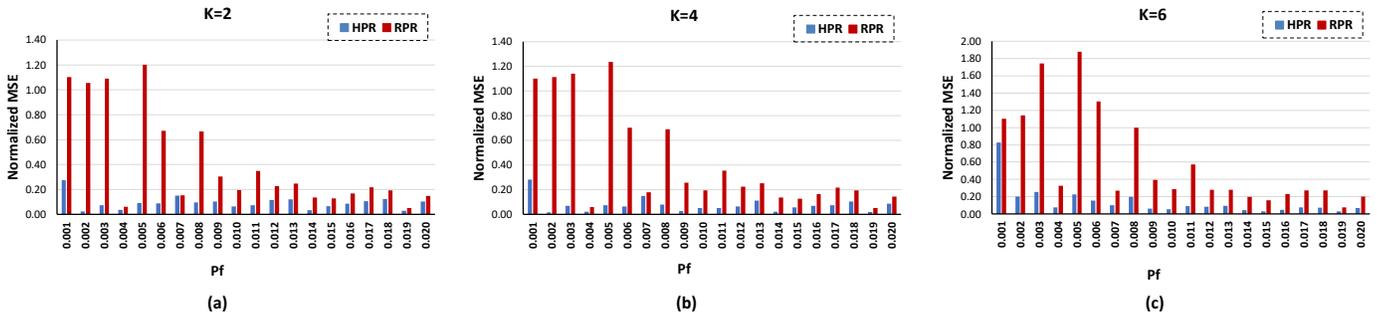

Fig. 9. MSE of the HPR and RPR multipliers normalized to the typical TMR multipliers, when the errors are injected into the internal signals of the studied designs with a uniform random distribution.

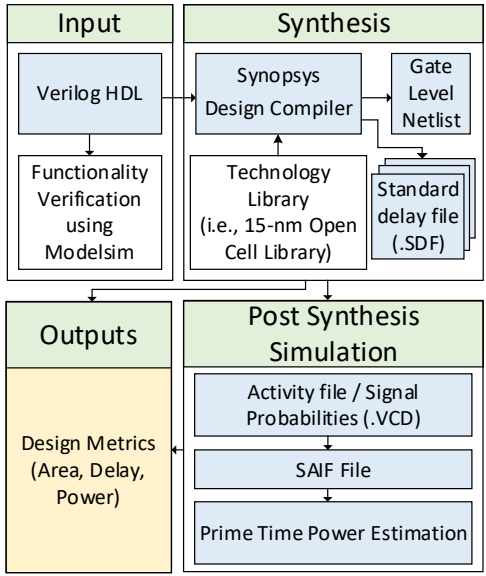

Fig. 10. The simulation setup and tool flow for synthesizing the investigated designs.

It can be seen that by increasing the value of $K$ the accuracy of the RPR scheme is decreased drastically (see the range of MSE in Fig. 8 a, b, and c), while for the HPR scheme, the MSE is significantly lower. These improvements are due to enhancing the accuracy of RP multipliers in the HPR structure, by exploiting intermediate signals of the FP multiplier (see Fig. 3). Note that the results in this figure were normalized to the typical TMR multiplier, e.g., at some points (e.g., see pointer ① in Figure 8.a) the normalized value of MSE is increased, which does not necessarily mean an increase in the amount of MSE.

We also examined the studied designs, when the errors are injected inside the designs. For this, first, we implemented all the studied designs at the gate level, such that their internal signals are available and we can inject errors into the internal signals. Fig. 9 shows the achieved results when the errors are injected into the internal signals of the designs. Based on the results, for $K$=2, 4, and 6, the proposed HPR schemes achieve, on average, 78%, 82%, and 77% lower MSE, respectively, compared to the state-of-the-art RPR multiplier. Note that compared to the case when the noise source is placed at the inputs of modules (i.e., Fig. 8), the results of Fig. 9 show the absolute variation of 1.9%, 0.7%, and 8.8%, respectively.

### B. Design Metrics

In this subsection, we compare the design metrics (i.e., area, delay, and power) of the proposed HPR scheme with those of the state-of-the-art RPR structure, as well as the typical TMR schemes. The simulation setup and used tool flow for these evaluations are shown in Fig. 10. As shown in this figure, the studied designs are implemented in Verilog HDL, and synthesized by the Synopsys Design Compiler (DC) using a 15nm Open Cell Library technology [33], with the normal operating voltage of 0.8V. Then, the Value Change Dump (VCD) file of the synthesized designs is created using the Modelsim HDL simulator, which corresponds to the activity of internal nodes. Note that the VCD file is obtained by injecting the real inputs (i.e., the studied image benchmarks) into the studied TMR designs. Then, the VCD files are translated to Switching Activity Interchange Format (SAIF) file that is used by Synopsys Prime Time to obtain the accurate power consumption of the synthesized designs.

Fig. 11 shows the design metrics of the studied designs obtained for the different values of $K$, ranging from 1 to 7. Based on the results, our proposed HPR multiplier achieved up to 69% and 70% lower area and power consumption, respectively, compared to the typical TMR-based multiplier, when $K$ is 7. However, the delay of the HPR is, on average, 5% higher than the typical TMR multiplier, for the different studied values of K. This overhead is due to extra signals in the HPR structure (see Fig. 4) compared to the typical TMR multiplier. In general, for the different values of $K$, the HPR multiplier offers, on average, 44% and 45% lower area and power consumption, compared to those of the typical TMR multiplier. Furthermore, the HPR multiplier led to slight overheads of, on average, 3% and 2% higher area and delay, respectively, when compared to the RPR multiplier, achieving a 3% lower delay. The extra delay of the RPR multiplier is due to the complex structure of its voter (see Fig. 1) compared to the simple voter of the proposed HPR multiplier (see Fig. 3).

### C. Image Processing Applications

In this subsection, we investigate the effectiveness of the proposed HPR multiplier in image processing applications, including image multiplication, image sharpening, and image smoothing filters, when given levels of noise are injected into the applications.

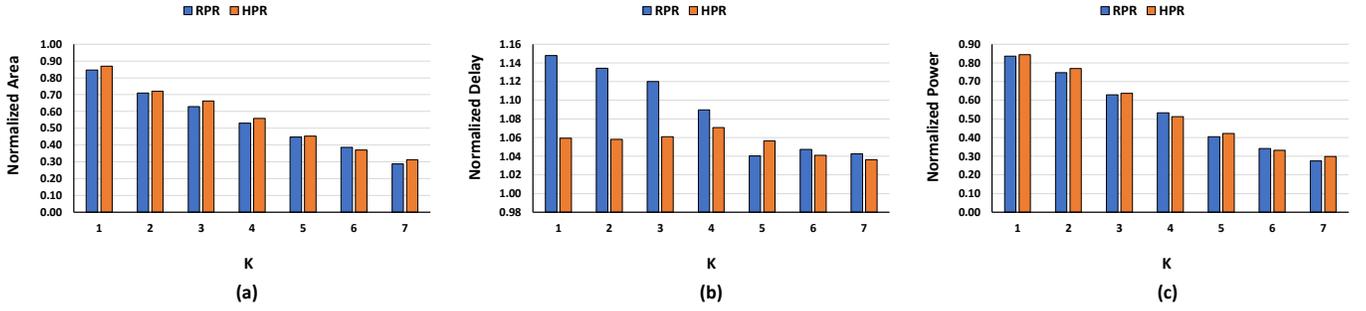

Fig. 11. Normalized a) area, b) delay, and c) power consumption of the studied redundancy-based multipliers compared to the typical TMR multiplier.

The selection of image processing for evaluating our proposed method in critical applications is due to getting the full potential of error-resilient approximable applications deployed in critical systems. For instance, as demonstrated in [12], cameras in an autonomous vehicle continuously receive images from the surrounding environment to extract important features for safety purposes, such as preventing lane departure. However, neither the human driver nor the autonomous system requires all the details of the images. Instead, the focus is on reliably extracting critical information from the images in the event of a system fault. Therefore, in this scenario, reliability takes precedence over the exact computation of unnecessary details. Our proposed HPR-Mul takes advantage of this opportunity to fully utilize approximate computing and address the challenges associated with typical TMR multiplier designs, such as their "strict majority" property and high sensitivity to soft errors.

For these evaluations, we leverage the noise injection model shown in Fig. 7. Also, in this work, six standard benchmark images from [35] are used. Note that in these image processing applications, similar to state-of-the-art works such as [11], and for simplicity, we consider a given value of K (i.e., K=4), which can be obtained from the quality degradation upper bound equations presented in equations (1) and (2). However, it would be of great value in this area if work is conducted proposing a systematic method for determining the quality level of approximate arithmetic of error-resilient applications deployed in critical systems, considering the reliability issues.

Furthermore, to investigate the quality of output images, the mean structural similarity index metric (MSSIM) is used. The MSSIM works based on the human visual systems that extract the information based on the structure of the image. MSSIM is calculated by [5]:

$$MSSIM(X,Y) = \frac{(2\mu_x\mu_y + C_1)(2\sigma_{xy} + C_2)}{(\mu_x^2 + \mu_y^2 + C_1)(\sigma_x^2 + \sigma_y^2 + C_2)} \quad (3)$$

where $\mu_x$, $\mu_y$, $\sigma_x$, $\sigma_y$, and $\sigma_{xy}$ are the local means, standard deviations, and cross-covariance of images $X$ and $Y$. Note that the MSSIM is in the range of 0 to 1, where the higher value of MSSIM corresponds to a higher quality.

TABLE II shows the MSSIM of output images obtained by the image multiplication application, for the two values of $P_f$ (i.e., 0.01 and 0.05), and when the parameter K of the RPR and HPR multipliers is 4. Based on the results, the image multiplication using the HPR multiplier achieves, on average, 7% (24%) and 9% (177%) higher MSSIM for the $P_f = 1\%$ and $P_f = 5\%$, respectively, compared to the RPR (Typical TMR)-based image multiplication. In the image sharpening application, the output image is obtained by [34]:

TABLE II  MSSIM OF IMAGE MULTIPLICATION USING THE DIFFERENT TMR SCHEMES, UNDER THE TWO VALUES OF $p_f$

| Benchmark Image | $P_f$ | Typical TMR | RPR | HPR |
|---|---|---|---|---|
| Lena × Sailboat on lake | 0.01 | 0.77 | 0.90 | 0.97 |
|  | 0.05 | 0.25 | 0.59 | 0.65 |
| Monarch Butterfly × Mandrill | 0.01 | 0.84 | 0.93 | 0.98 |
|  | 0.05 | 0.28 | 0.69 | 0.74 |
| Female × Tree | 0.01 | 0.75 | 0.90 | 0.97 |
|  | 0.05 | 0.21 | 0.60 | 0.66 |
| **Average** | **0.01** | **0.79** | **0.91** | **0.97** |
|  | **0.05** | **0.25** | **0.63** | **0.68** |

$$Y(i,j) = 2X(i,j) - \frac{1}{273}\sum_{m=-2}^{2}\sum_{n=-2}^{2} X(i+m, j+n) \cdot \text{Mask}_{\text{Sharpening}}(m+3, n+3) \quad (4)$$

where $X$ and $Y$ are the input and output images, respectively. Also, Mask$_{\text{Sharpening}}$ matrix is

$$\text{Mask}_{\text{Sharpening}} = \begin{bmatrix} 1 & 4 & 7 & 4 & 1 \\ 4 & 16 & 26 & 16 & 4 \\ 7 & 26 & 41 & 26 & 7 \\ 4 & 16 & 26 & 16 & 4 \\ 1 & 4 & 7 & 4 & 1 \end{bmatrix} \quad (5)$$

In TABLE III, the MSSIM of output images of the image sharpening filter under the different values of $P_f$ is presented. Based on the results, the HPR-based image sharpening filter achieves, on average, 11% (32%) and 13% (49%) higher MSSIM for the $P_f = 1\%$ and $P_f = 5\%$, respectively, compared to the RPR (Typical TMR)-based image sharpening filter.

To achieve the smoothed output image, the following equation is used [34]:

$$Y(i,j) = \frac{1}{60}\sum_{m=-2}^{2}\sum_{n=-2}^{2} X(i+m, j+n) \cdot \text{Mask}_{\text{Smoothing}}(m+3, n+3) \quad (6)$$

where $X$ and $Y$ are the input and output images, respectively, and Mask$_{\text{Smoothing}}$ matrix is realized by:

$$\text{Mask}_{\text{Smoothing}} = \begin{bmatrix} 1 & 1 & 1 & 1 & 1 \\ 1 & 4 & 4 & 4 & 1 \\ 1 & 4 & 12 & 4 & 1 \\ 1 & 4 & 4 & 4 & 1 \\ 1 & 1 & 1 & 1 & 1 \end{bmatrix} \quad (7)$$

TABLE III  MSSIM OF IMAGE SHARPENING USING THE DIFFERENT TMR SCHEMES, FOR THE TWO VALUES OF $p_f$

| Benchmark Image | $P_f$ | Typ. | RPR | HPR |
|---|---|---|---|---|
| Sailboat on lake | 0.01 | 0.747 | 0.885 | 0.986 |
|  | 0.05 | 0.607 | 0.823 | 0.929 |
| Lena | 0.01 | 0.793 | 0.892 | 0.992 |
|  | 0.05 | 0.670 | 0.855 | 0.960 |
| Monarch Butterfly | 0.01 | 0.784 | 0.884 | 0.985 |
|  | 0.05 | 0.683 | 0.824 | 0.931 |
| Mandrill | 0.01 | 0.725 | 0.884 | 0.985 |
|  | 0.05 | 0.617 | 0.825 | 0.929 |
| Tree | 0.01 | 0.737 | 0.877 | 0.980 |
|  | 0.05 | 0.603 | 0.813 | 0.921 |
| Female | 0.01 | 0.687 | 0.878 | 0.979 |
|  | 0.05 | 0.568 | 0.797 | 0.905 |
| **Average** | **0.01** | **0.746** | **0.883** | **0.985** |
|  | **0.05** | **0.625** | **0.823** | **0.929** |

The MSSIM of output images obtained by the image smoothing filter application, are shown in TABLE IV. Based on the results, the image smoothing filter using the HPR multiplier results in, on average, 11% (212%) and 11% (1065%) higher MSSIM for the $P_f = 1\%$ and $P_f = 5\%$, respectively, compared to the RPR (Typical TMR)-based image smoothing filter. Note that the output images of the typical TMR-based image smoothing filter have a very low and almost unacceptable quality, while the HPR-based filter offers the highest quality.

Fig. 12 shows the input and output images of the image sharpening and smoothing filters leveraging the different studied redundancy-based multipliers for the "*Monarch Butterfly*" benchmark image, under the two $P_f$ values of 0.01 and 0.05. As the results of TABLE IV and Fig. 12 show, the typical TMR-based image smoothing filter has a very low quality compared to the other studied image processing applications.

This behavior is because the image smoothing filter generates output pixels with a lower value compared to the other studied image processing applications. In detail, the range of matrix elements (5) is between 1 and 41, while this is between 1 and 12 in (7), i.e., in the image smoothing filter, the pixels of input images are multiplied in numbers with a significantly lower range, which results in lower output value rather than other studied image processing applications. Moreover, the typical TMR designs are sensitive to a slight difference between the module's output, which is referred to as the "strict majority" property [5]. Therefore, the typical TMR-based image smoothing filter has an almost unacceptable quality compared to the other image processing applications, in the presence of errors.

TABLE IV  MSSIM OF THE IMAGE SMOOTHING FILTER USING THE DIFFERENT TMR SCHEMES, FOR THE TWO VALUES OF $p_f$

| Benchmark Image | $P_f$ | Typ. | RPR | HPR |
|---|---|---|---|---|
| Sailboat on lake | 0.01 | 0.333 | 0.898 | 0.997 |
|  | 0.05 | 0.092 | 0.876 | 0.976 |
| Lena | 0.01 | 0.370 | 0.899 | 0.999 |
|  | 0.05 | 0.095 | 0.892 | 0.992 |
| Monarch Butterfly | 0.01 | 0.335 | 0.899 | 0.999 |
|  | 0.05 | 0.089 | 0.888 | 0.988 |
| Mandrill | 0.01 | 0.298 | 0.898 | 0.998 |
|  | 0.05 | 0.068 | 0.883 | 0.983 |
| Tree | 0.01 | 0.260 | 0.898 | 0.998 |
|  | 0.05 | 0.060 | 0.884 | 0.985 |
| Female | 0.01 | 0.326 | 0.898 | 0.998 |
|  | 0.05 | 0.103 | 0.886 | 0.985 |
| **Average** | **0.01** | **0.320** | **0.898** | **0.998** |
|  | **0.05** | **0.085** | **0.885** | **0.985** |

| | | | Image Sharpening | | | | |
|---|---|---|---|---|---|---|---|
| | | | $P_f = 0.01$ | | | $P_f = 0.05$ | |
| | Error-free sharpened image | Typical TMR Multiplier | RPR | HPR | Typical TMR Multiplier | RPR | HPR |
| Image | 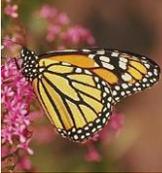 | 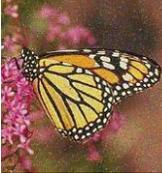 | 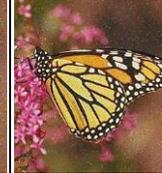 | 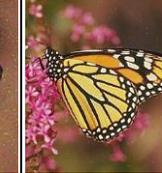 | 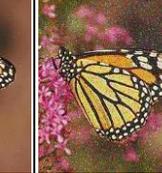 | 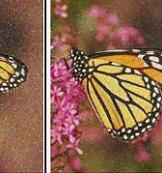 | 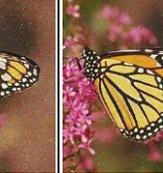 |
| **MSSIM** | 1.000 | 0.784 | 0.884 | 0.985 | 0.683 | 0.824 | 0.931 |
| | | | Image Smoothing | | | | |
| | Error-free smoothed image | Typical TMR Multiplier | RPR | HPR | Typical TMR Multiplier | RPR | HPR |
| | | | $P_f = 0.01$ | | | $P_f = 0.05$ | |
| Image | 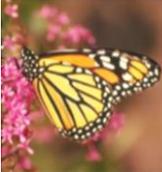 | 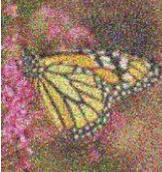 | 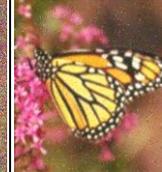 | 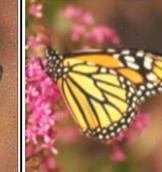 | 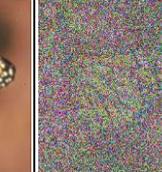 | 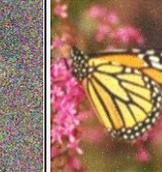 | 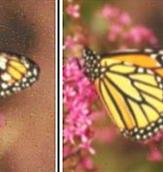 |
| **MSSIM** | 1.000 | 0.335 | 0.899 | 0.999 | 0.089 | 0.888 | 0.988 |

Fig. 12. Input and output images of the image sharpening and smoothing filters leveraging the different studied redundancy-based multipliers, for the "*Monarch Butterfly*" benchmark image, under the two $P_f$ values of 0.01 and 0.05.

Finally, Fig. 13 shows the power consumption of the image multiplication application using the HPR-Mul scheme for the three different values of $K$ (i.e., $K$=2, 4, and 6), in which the power results were normalized to the typical TMR multiplier. Based on the results shown in this figure, for the Female × Tree, Lena × Sailboat on Lake, and Monarch Butterfly × Mandril image multiplications, the proposed HPR-Mul achieved, on average, 42%, 47%, and 37% lower power consumption compared to the typical TMR multiplier, respectively, for the three studied values of K.

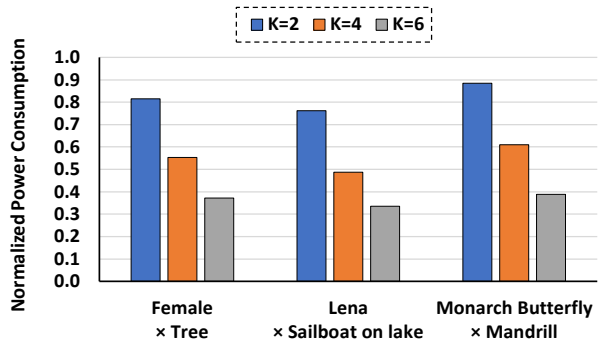

Fig. 13. Normalized power consumption of the HPR-Mul for the image multiplication application under the various values of $K$.

## V. Conclusion

In this paper, we presented a high precision redundancy multiplier (HPR-Mul) structure that leverages a full precision multiplier along with the two reduced precision multipliers, where the output is determined by a simple majority-based voter. Employing these reduced precision modules provides a higher energy efficiency and lower area while resolving the challenges of the typical TMR schemes, such as the high sensitivity to soft errors and the design diversity of the triplicated module. Unlike prior state-of-the-art reduced precision redundancy multipliers (RPR-Mul), the proposed HPR-Mul employs some signals of the full precision multiplier to enhance the accuracy of the reduced precision ones resulting in a considerably higher accuracy. The HPR-Mul showed up to 69% and 70% lower area and power consumption compared to the typical TMR multiplier, at the cost of 4% higher delay. Moreover, the HPR-Mul outperforms the RPR-Mul by achieving up to 84% higher soft error tolerance. Finally, in the examined image processing applications, the HPR-based applications showed up to 13% (1065%) higher MSSIM in comparison with the state-of-the-art RPR (typical TMR-based) multipliers, respectively.

## References

[1] A. A. Bahoo, O. Akbari and M. Shafique, "An Energy-Efficient Generic Accuracy Configurable Multiplier Based on Block-Level Voltage Overscaling," in *IEEE Transactions on Emerging Topics in Computing*, vol. 11, no. 4, pp. 851-867, Oct.-Dec. 2023.

[2] I. Koren and C. Krishna, *Fault-Tolerant Systems.*, Morgan Kaufmann, 2007.

[3] I. Alzaher-Noufal and M. Nicolaidis, "A CAD Framework for Generating Self-Checking Multipliers Based on Residue Codes", in Proceedings of the Conference on Design, Automation and Test in Europe (DATE), 1999.

[4] M. Nicolaidis and R. O. Duarte, "Fault-Secure Parity Prediction Booth Multipliers", IEEE Design & Test of Computers, vol. 16, no. 3, pp. 90-101, 1999.

[5] J. Vafaei, O. Akbari, M. Shafique and C. Hochberger, "X-Rel: Energy-Efficient and Low-Overhead Approximate Reliability Framework for Error-Tolerant Applications Deployed in Critical Systems," in *IEEE Transactions on Very Large Scale Integration (VLSI) Systems*, vol. 31, no. 7, pp. 1051-1064, July 2023.

[6] D. Kobayashi, "Scaling Trends of Digital Single-Event Effects: A Survey of SEU and SET Parameters and Comparison With Transistor Performance," in *IEEE Transactions on Nuclear Science*, vol. 68, no. 2, pp. 124-148, Feb. 2021.

[7] S. Byonghyo, S. R. Sridhara, and N. R. Shanbhag, "Reliable low-power digital signal processing via reduced precision redundancy," IEEE Transac-tions on Very Large Scale Integration (VLSI) Systems, vol. 12, no. 5, pp. 497-510, 2004.

[8] B. Shim and N. R. Shanbhag, "Energy-efficient soft error-tolerant dig-ital signal processing," in IEEE Transactions on Very Large Scale Integration (VLSI) Systems, vol. 14, no. 4, pp. 336-348, 2006.

[9] S. LIU, K. Chen, P. Reviriego, W. Liu, A. LOURI and F. Lombardi, "Reduced Precision Redundancy for Reliable Processing of Data," in *IEEE Transactions on Emerging Topics in Computing*, vol. 9, no. 4, pp. 1960-1971, 1 Oct.-Dec. 2021.

[10] A. Ullah, P. Reviriego, S. Pontarelli and J. A. Maestro, "Majority Voting-Based Reduced Precision Redundancy Adders," *IEEE Trans. Device and Materials Reliability,*, vol. 18, no. 1, pp. 122-124, Mar. 2018.

[11] K. Chen, L. Chen, P. Reviriego and F. Lombardi, "Efficient Implementations of Reduced Precision Redundancy (RPR) Multiply and Accumulate (MAC)," in *IEEE Transactions on Computers*, vol. 68, no. 5, pp. 784-790, 1 May 2019.

[12] F. Baharvand and S. G. Miremadi, "LEXACT: Low Energy N-Modular Redundancy Using Approximate Computing for Real-Time Multicore Processors", *IEEE Trans. Emerging Topics in Computing*, pp. 1-1, 2017.

[13] W. Liu, F. Lombardi and M. Shulte, "A Retrospective and Prospective View of Approximate Computing," in *Proceedings of the IEEE*, vol. 108, no. 3, pp. 394-399, March 2020.

[14] G. Rodrigues, et al., "Survey on Approximate Computing and Its Intrinsic Fault Tolerance," in *Electronics*, vol. 9, no. 4, p. 557, Mar. 2020.

[15] A. Aponte-Moreno, A. Moncada, F. Restrepo-Calle and C. Pedraza, "A review of approximate computing techniques towards fault mitigation in HW/SW systems," *IEEE 19th Latin-American Test Symposium (LATS)*, Brazil, 2018, pp. 1-6.

[16] A. Bosio, I. O'Connor, G. S. Rodrigues, F. K. Lima and S. Hamdioui, "Exploiting Approximate Computing for implementing Low Cost Fault Tolerance Mechanisms," *15th Design & Technology of Integrated Systems in Nanoscale Era (DTIS)*, Morocco, 2020, pp. 1-2.

[17] K. Chen, F. Lombardi and J. Han, "An approximate voting scheme for reliable computing," *Design, Automation & Test in Europe Conference & Exhibition (DATE)*, 2015, pp. 293-296.

[18] K. Chen, J. Han and F. Lombardi, "Two Approximate Voting Schemes for Reliable Computing," in *IEEE Transactions on Computers*, vol. 66, no. 7, pp. 1227-1239, 1 July 2017.

[19] G.S. Rodrigues, et al., "Approximate TMR based on successive approximation and loop perforation in microprocessors," in *Microelectronics Reliability*, vol. 100–101, sep. 2019.

[20] I. Polian and J. P. Hayes, "Selective Hardening: Toward Cost-Effective Error Tolerance," *IEEE Design & Test of Computers*, vol. 28, no. 3, pp. 54-63, May 2011.

[21] Iuri A.C. Gomes, et al., "Exploring the use of approximate TMR to mask transient faults in logic with low area overhead," in *Microelectronics Reliability*, Vol. 55, no. 9–10, pp. 2072-2076, 2015.

[22] A. J. Sánchez-Clemente, L. Entrena and M. García-Valderas, "Partial TMR in FPGAs Using Approximate Logic Circuits," *IEEE Trans. Nuclear Science*, vol. 63, no. 4, pp. 2233-2240, Aug. 2016.

[23] T. Arifeen, et al., "A Fault Tolerant Voter for Approximate Triple Modular Redundancy," in *Electronics*, vol. 8, no. 3, pp. 332, Mar. 2019.

[24] B. Pratt, M. Fuller, M. Rice, and M. Wirthlin, "Reduced-precision redundancy for reliable FPGA communications systems in high-radiation environments," *IEEE Trans. Aerosp. Electron. Syst.*, vol. 49, no. 1, pp. 369–380, Jan. 2013.

[25] M. A. Sullivan, "Reduced precision redundancy applied to arithmetic operations in field programmable gate arrays for satellite control and sensor systems," M.S. thesis, Dept. Elect. Comput. Eng., Naval Postgraduate School, Monterey, CA, USA, 2008.

[26] A. Aponte-Moreno, C. Pedraza and F. Restrepo-Calle, "Reducing Overheads in Software-based Fault Tolerant Systems using Approximate Computing," *IEEE Latin American Test Symposium (LATS)*, Chile, 2019, pp. 1-6.

[27] M. Biasielli, C. Bolchini, L. Cassano, A. Mazzeo and A. Miele, "Approximation-Based Fault Tolerance in Image Processing Applications," *in IEEE Transactions on Emerging Topics in Computing*, vol. 10, no. 2, pp. 648-661, 1 April-June 2022.

[28] A. Aponte-Moreno, F. Restrepo-Calle and C. Pedraza, "FTxAC: Leveraging the Approximate Computing Paradigm in the Design of Fault-Tolerant Embedded Systems to Reduce Overheads," in *IEEE Transactions on Emerging Topics in Computing*, vol. 9, no. 2, pp. 797-810, 1 April-June 2021.

[29] O. Akbari, M. Kamal, A. Afzali-Kusha, M. Pedram and M. Shafique, "X-CGRA: An Energy-Efficient Approximate Coarse-Grained Reconfigurable Architecture," in *IEEE Transactions on Computer-Aided Design of Integrated Circuits and Systems*, vol. 39, no. 10, pp. 2558-2571, Oct. 2020.

[30] O. Akbari, et al., "RAP-CLA: A Reconfigurable Approximate Carry Look-Ahead Adder," in *IEEE Transactions on Circuits and Systems II: Express Briefs*, preprint, 23 May. 2017.

[31] S. S. Venkataramani, et al., "Quality programmable vector processors for approximate computing," in *Proc. 46th Annual International Symposium on Microarchitecture (MICRO)*, Dec. 2013, pp. 1–12.

[32] A.S. Hassan, et al., "Generation Methodology for Good-Enough Approximate Modules of ATMR", in *Journal of Electronic Testing*, vol. 34, pp. 651–665, 2018.

[33] M. Martins et al., "Open cell library in 15nm FreePDK technology," in *Proc. Symp. Int. Symp. Phys. Design (ISPD)*, 2015, pp. 171–178.

[34] O. Akbari, M. Kamal, A. Afzali-Kusha, and M. Pedram, "Dual Quality 4:2 Compressors for Utilizing in Dynamic Accuracy Configurable Multipliers," *IEEE Trans. Very Large Scale Integr. (VLSI) Syst.*, vol. 25, no. 4, pp. 1352–1361, Apr. 2017.

[35] *The USC-SIPI Image Database*. Accessed: Mar. 16, 2024. [Online]. Available: http://sipi.usc.edu/database/

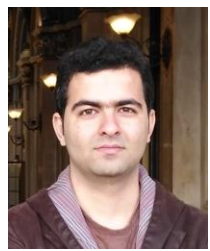

**Omid Akbari** received the B.Sc. degree from the University of Guilan, Rasht, Iran, in 2011, the M.Sc. degree from Iran University of Science and Technology, Tehran, Iran, in 2013, and the Ph.D. degree from the University of Tehran, Iran, in 2018, all in Electrical Engineering, Electronics - Digital Systems sub-discipline. He was a visiting researcher in the CARE-Tech Lab. at Vienna University of Technology (TU Wien), Austria, from Apr. to Oct. 2017, and a visiting research fellow under the Future Talent Guest Stay program at Technische Universität Darmstadt (TU Darmstadt), Germany, from Jul. to Sep. 2022. He is currently an assistant professor of Electrical and Computer Engineering at Tarbiat Modares University, Tehran, Iran, where he is also the Director of the Computer Architecture and Dependable Systems Laboratory (CADS-Lab). His current research interests include embedded machine learning, reconfigurable computing, energy-efficient computing, distributed learning, and fault-tolerant system design.

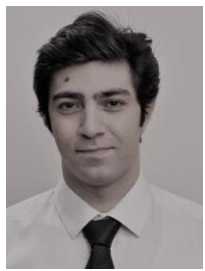

**Jafar Vafaei** received the B.Sc. degree from the University of Tabriz, Tabriz, Iran, in 2013, and the M.Sc. degree from University of Tehran, Tehran, Iran, in 2019, both in Electrical Engineering. He is currently a research assistant at Tarbiat Modares University, Tehran, Iran in the Computer Architecture and Dependable systems Laboratory (CADS-Lab). His research interests include low power digital designs and machine learning, reconfigurable computing, neuromorphic computing, and fault-tolerant system design.